\documentclass[prb,twocolumn,showpacs]{revtex4}
\usepackage{amsfonts} 
\usepackage{amsmath}
\usepackage{amssymb}
\usepackage{bm}
\usepackage[usenames, dvipsnames]{color} 
\usepackage{dcolumn}
\usepackage{epic} 
\usepackage{epsfig}
\usepackage{eucal} 
\usepackage{graphicx}
\usepackage[breaklinks,colorlinks = true,linkcolor = red,urlcolor=Magenta,citecolor=red]{hyperref}
\usepackage{mathrsfs}
\usepackage{soul}
\usepackage{subfigure}
\usepackage{wrapfig} 
\usepackage{xy} 

\begin{document}
\title{Fractionalized topological insulators from frustrated spin models in three dimensions}

\author{Subhro Bhattacharjee$^{1,2}$}\email{subhro@physics.utoronto.ca}
\author{Yong Baek Kim$^{1,3}$}
\author{Sung-Sik Lee$^{2,4}$}
\author{Dung-Hai Lee$^{5,6}$}
\affiliation{$^1$ Department of Physics, University of Toronto, Toronto, Ontario M5S 1A7, Canada.\\
$^2$ Department of Physics and Astronomy, McMaster University, Hamilton, Ontario  L8S 4M1, Canada.\\
$^3$ School of Physics, Korea Institute for Advanced Study, Seoul 130-722, Korea.\\
$^4$ Perimeter Institute for Theoretical Physics, Waterloo ON N2L 2Y5, Canada.\\
$^5$ Department of Physics,University of California at Berkeley, Berkeley, CA 94720, USA.\\
$^6$ Materials Sciences Division, Lawrence Berkeley National Laboratory, Berkeley, CA 94720, USA.}

\date{\today}
\begin{abstract}
We present the theory of a three-dimensional fractionalized topological insulator in the form of a $U(1)$ spin liquid with gapped fermionic spinons in the bulk and topologically protected gapless spinon surface states. Starting from a spin-rotation-invariant spin-$1/2$ model on a pyrochlore lattice with frustrated antiferromagnetic and ferromagnetic exchange interactions, we show that decomposition of the latter interactions, within slave-fermion representation of the spins, can naturally give rise to an emergent spin-orbit coupling for the spinons by spontaneously breaking the spin rotation symmetry. The time-reversal symmetry, however, is preserved. This stabilizes a fractionalized topological insulator which also has bulk bond spin-nematic order. We describe the low-energy properties of this state.

\end{abstract}

\maketitle
\section{Introduction}
A central question in the recent attempts {\cite{2011_chen}} to classify  topological order in gapped quantum states is whether there exists a sequence of local unitary transformations connecting a given ground state wave function to a trivial product wave function, without closing the bulk energy gap. If such a ``path" can be found then the given state is adiabatically connected to a trivial gapped state, and hence devoid of any topological order. 

For a class of systems, the ground-state wave functions cannot be deformed, through {\em any} sequence of such transformations, into a trivial product state, unless the bulk gap is closed. Gapped quantum spin liquids {\cite{2002_wen}} and quantum Hall states {\cite{1997_dassarma}} are particular examples of this kind of topological order. Such topological order is a signature of underlying long-range quantum entanglement. \cite{2011_chen} In another class, there are systems where such paths can be found {\it only} if transformations violating certain symmetries such as time-reversal/particle-hole transformations are allowed. Once such ``symmetry violating" paths are excluded, this second class of systems is no longer adiabatically connected to the trivial atomic insulator and hence exhibits a kind of ``symmetry protected topological order." {\cite{2011_chen}} Examples include strong topological band insulators and topological superconductors. {\cite{2010_hasan,2007_fu,2010_fu}} In both the above cases, the systems may or may not posses further discrete/continuous global symmetries such as lattice translation or spin rotation.

In this paper, we introduce a three-dimensional fractionalized topological insulator (TI) in the form of a $U(1)$ spin liquid which also breaks spin rotation symmetry spontaneously. Here, we have broadened the usual definition of a spin liquid (as quantum paramagnets without any broken symmetry \cite{2002_wen}) to include  all states having deconfined fractionalized spinon excitations that are minimally coupled to an emergent gauge field. This definition encompasses both symmetric, \cite{2002_wen} and symmetry-broken spin liquids. \cite{1999_balents,2000_senthil,2004_senthil} The state considered here has gapped fermionic spinons as well-defined low-energy quasiparticles and gapless ``photons" of an emergent compact $U(1)$ gauge field. \cite{2002_wen} It also exhibits bond spin-nematic order in the bulk. {\cite{2009_shindou}} Due to this spin-rotation breaking, there are three Goldstone modes in the low-energy spectrum.  However, this spin liquid is different from the usual broken-symmetry states where the ground-state wave function may be deformed to a trivial product state. In the case of this spin liquid, the presence of the emergent gauge field, minimally coupled with the fractionalized excitations (spinons), leads to robust long-range entanglement among the underlying spins. Side by side, since the low-energy effective Hamiltonian is time-reversal invariant (the spin-nematic order does not break time reversal symmetry), the spinons have an emergent topologically non trivial band structure;  i.e., the spinon ``bands" carry a non-zero $Z_2$ index similar to topological band insulators in non interacting electronic systems. {\cite{2007_fu,2010_hasan}} In this sense the present state exhibits topological order of the second kind.

Usually, the notion of topological order is reserved for systems where all bulk excitations are separated from the ground state by an energy gap. In the present case, however, while the spinons are gapped, there are gapless photons and Goldstone bosons in the bulk. While it is easy to gap out the Goldstone bosons by explicitly breaking the spin rotation symmetry of the Hamiltonian (by including small spin anisotropy), the photons are rather robust and cannot be gapped out without actually destroying the $U(1)$ spin liquid. This is because, the presence of the photon is a direct consequence of the fractionalization of the electrons and the stability of the spin liquid itself guarantees the existence of the photon. \cite{2002_wen} To move out of this spin liquid phase, we need to close the spinon gap and/or confine the spinons (monopole condensation, see below). Hence, as long as the spinons are present and the spinon gap is well defined throughout the Birllouin zone, we find it plausible to consider the extension of concepts of symmetry-protected topological order along with the long-range entanglement to the present case. Indeed, the state under consideration can {\em only} be described fruitfully if we take into account all three aspects of modern theory of phases of condensed matter : fractionalization, broken symmetry, and topological order.
 
In the following, we construct a fractionalized topological insulator on the pyrochlore lattice within self consistent slave-particle mean-field theory for a $SU(2)$ symmetric Heisenberg model. We start with the description of the spin model on a pyrochlore lattice in Section \ref{sec_spin_model} and argue that this model has a chance of stabilizing a spin liquid ground state. The particular spin liquid that we consider has a non trivial topology of the spinon band structure that emerges as a consequence of many-body correlations and is not inherited from that of the underlying electrons. We show that the slave fermion decomposition of ferromagnetic spin-spin interactions in the triplet sector {\cite{2009_shindou}} naturally gives rise to an effective ``spin-orbit" (SO) coupling for the spinons [Eq. (\ref{eq_fm_decouple})]. This explicit construction, as discussed in Section \ref{sec_ansatz}, of an emergent SO coupling contrasts with the recent attempts to obtain such topological Mott insulators. \cite{2008_young,2009_levin,2011_levin,2010_pesin,2010_krempa,2010_rachel,2011_swingle} These works use the parton construction, where the electron is fragmented into a number of partons and the topological band structure of the partons are either inherited from the underlying electrons or are assumed to exist for the partons. We discuss the unusually rich low-energy spectrum of the state, its surface states, and comment on beyond mean field effects as well as possible phase transitions out of it in Section \ref{sec_excitation}. The details of the calculations are discussed in the appendices.


\section{The spin model}
\label{sec_spin_model}
\begin{figure}
\centering
\includegraphics[scale=0.4]{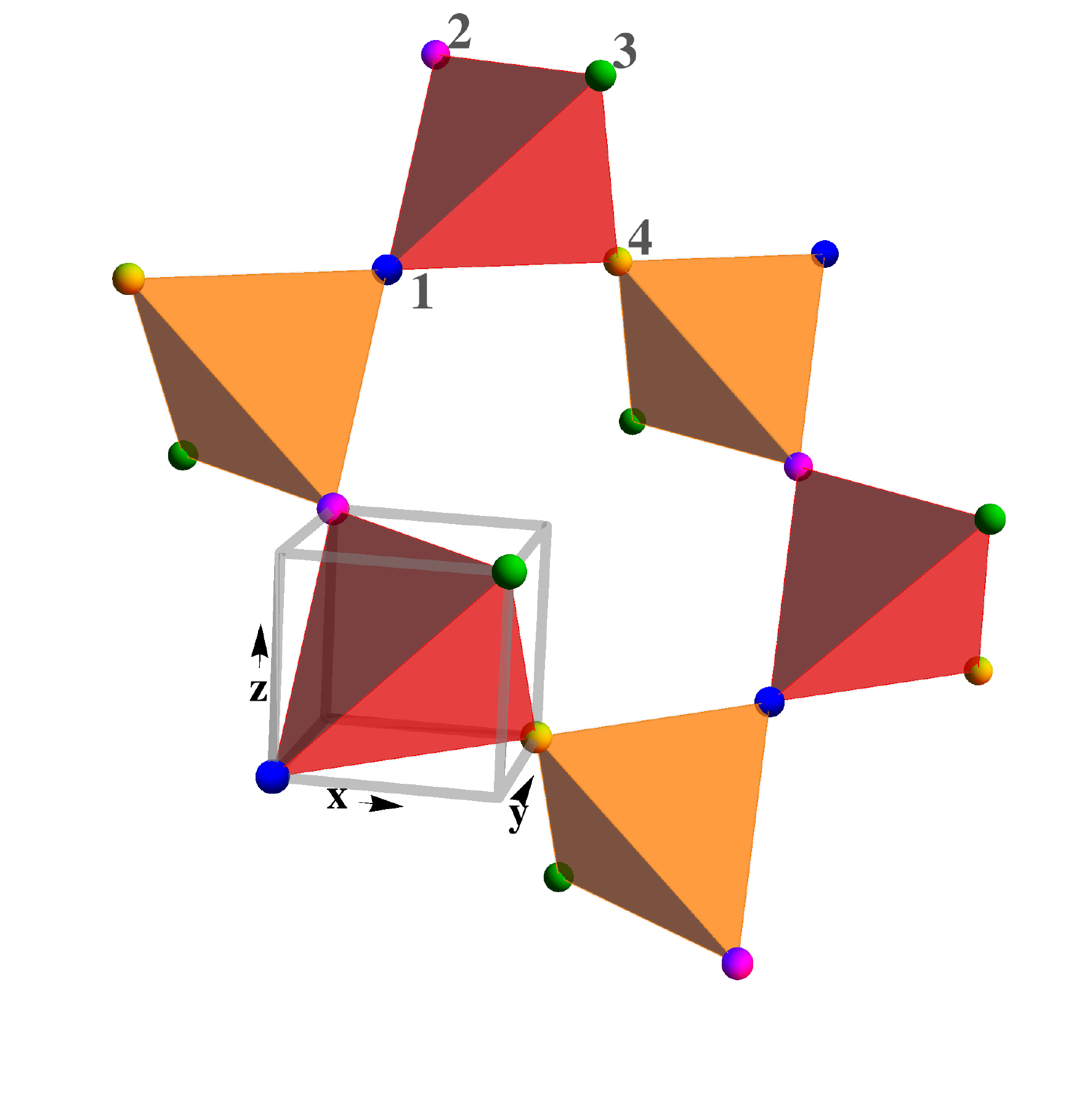}
\caption{The pyrochlore lattice can be described as a fcc lattice with a 4-point basis. The 4 sublattices are numbered as $a=1,2,3,4$. We have used the notation of Ref. {\onlinecite{2010_conlon}}.}
\label{fig_pyrochlore}
\end{figure}

 It is known that spin systems on geometrically frustrated lattices such as the pyrochlore are a good place to look for spin liquids. The nearest-neighbour Heisenberg antiferromagnet on this lattice is highly frustrated, even at the classical level. \cite{2004_isakov} For lower spin values (including $S=1/2$), where quantum fluctuations are generally known to suppress magnetic order, a three-dimensional spin liquid ground state is expected (Ref. \onlinecite{2008_burnell} and references therein). Further, the pyrochlore lattice has a 4-sublattice structure (see Fig. {\ref{fig_pyrochlore}}); i.e., it has four sites per unit cell. Since for spin-$1/2$ there is one spinon per site, this lattice can in principle support a gapped $U(1)$ spin liquid in the form of a {``spinon band insulator.''} In passing, we note that such a gapped $U(1)$ spin liquid state, in two spatial dimensions, is unstable to confinement of the spinons.\cite{1987_polyakov}

Consider the extended Heisenberg model on the pyrochlore lattice (shown in Fig. {\ref{fig_pyrochlore}}). 
\begin{align}
H=J_1\sum_{\langle ij \rangle} {\bf S}_i\cdot{\bf S}_j -J_2\sum_{\langle\langle ij\rangle\rangle}{\bf S}_i\cdot{\bf S}_j\label{eq_hamiltonian}
\end{align}
where ${\bf S}_i$ are spin-$1/2$ operators at the site $i$; $\langle ij\rangle$ and $\langle\langle ij\rangle\rangle$  indicate sum over the first and second-nearest neighbours respectively (details of the pyrochlore lattice are discussed in Appendix \ref{appen_pyro}) and $J_1,J_2>0$. This indicates that nearest and second-neighbour interactions are antiferromagnetic and ferromagnetic, respectively. For the classical model, in the absence of $J_2$, the ground state is extensively degenerate (see Ref. {\onlinecite{2004_isakov}} and references therein). On incorporating $J_2$ it shows magnetic ordering. {\cite{1991_reimers}} However, such a magnetic order is rather weak when $J_2/J_1\ll 1$. In such a regime, the corresponding model for the $S=1/2$ case may stabilize a spin liquid ground state. 

In this paper, we consider a particular class of spin liquids with non-collinear bond spin-nematic order in the bulk. Our calculations show that such spin liquids indeed represent self-consistent saddle point solutions for the above Hamiltonian with reasonably competitive ground-state energies at the mean-field level. We therefore argue that such spin liquids are indeed good candidates  for the ground state for the above and related models. 
\section{Mean field theory for the spinon topological insulator} 
\label{sec_ansatz}

In this section, we consider the possible paramagnetic ground states of the spin model [Eq. \ref{eq_hamiltonian}] within mean-field theory.
\subsection{The mean-field {\em Ansatz}}
In magnetically disordered spin liquids, we have $\langle {\bf S}_i\rangle=0$. For ferromagnetic spin exchange, the spinon decoupling in the triplet channel is favoured. This is given by {\cite{2009_shindou}}

\begin{align}
\nonumber
-{\bf {S}}_i\cdot{\bf{S}}_j=&
\frac{3}{8}\left(\vert {\bf{E}}_{ij}\vert^2+\vert{\bf D}_{ij}\vert^2\right)+const\\
\nonumber
-&\frac{3}{8}\left(E^*_{ij,a}f^\dagger_{i\alpha}\sigma^a_{\alpha\beta}f_{j\beta}+h.c.\right)\\
-&\frac{3}{8}\left(D^*_{ij,a}f_{i\alpha}\left[\imath\sigma^a\sigma^2\right]_{\alpha\beta}f_{j\beta} +h.c.\right)
\label{eq_fm_decouple}
\end{align}
where $f_{i\alpha}$ is the fermionic spinon annihilation operator at site $i$ and spin $\alpha$ and
\begin{align}
E_{ij,a}=\langle f^\dagger_{i,\alpha}\sigma^a_{\alpha\beta}f_{j\beta}\rangle, \\
D_{ij,a}=\langle f_{i\alpha}\left[\imath\sigma^a\sigma^y\right]_{\alpha\beta}f_{j\beta}\rangle
\end{align}
($a=x,y,z$ and $\sigma^a$ are the Pauli matrices) are mean fields corresponding to the triplet particle-hole and particle-particle channels respectively for the spinons. Thus, in the above spin Hamiltonian [Eq. \ref{eq_hamiltonian}], the second neighbours, which are ferromagnetic, must be decoupled in this triplet channel. 

This is in addition to the usual decoupling of the antiferromagnetic exchanges along the singlet channel (for nearest neighbours), {\cite{1992_ubbens}}
\begin{align}
\nonumber
{\bf {S}}_i\cdot{\bf{S}}_j=&
\frac{3}{8}\left(\vert \chi_{ij}\vert^2+\vert\Delta_{ij}\vert^2\right)+constant\\
\nonumber
-&\frac{3}{8}\left(\chi^*_{ij}  f^{\dagger}_{i\alpha}f_{j\alpha}+h.c.\right)\\
-&\frac{3}{8}\left(\Delta^*_{ij}f_{i\alpha}\left[i\sigma^2\right]_{\alpha\beta}f_{j\beta}+h.c.\right)
\label{eq_afm_decouple}
\end{align}
where, 
\begin{align}
\chi_{ij}=\langle f^\dagger_{i\alpha}f_{j\alpha}\rangle\\
\Delta_{ij}=\langle f_{i\alpha}\left[i\sigma^y\right]_{\alpha\beta}f_{j\beta}\rangle.
\end{align}  

 In the present case, we choose spin liquid {\em Ans\"{a}tze} that only consider the particle-hole channel and all particle-particle pairing channels are set to zero. This is equivalent to setting $\Delta_{ij},{\bf D}_{ij}=0$ identically. The presence of the particle-particle channels would reduce the gauge group (see below and also Ref. \onlinecite{2002_wen}) from $U(1)$ to $Z_2$.  With such pairings one may be able to stabilize a gapped $Z_2$ spin liquid with topological spinon band structure. However, we did not find a simple and stable $Z_2$ spin liquid {\em Ansatz}  for the present model. Therefore, for the rest of this paper we shall concentrate on $U(1)$ spin liquids.

The spinon spectrum then depends on the structure of ${\bf E}_{ij}$ and $\chi_{ij}$. The above ansatz is invariant under the $U(1)$ gauge transformation of the spinons: $f_{i\sigma}\rightarrow e^{i\phi_i}f_{i\sigma}$. Therefore within the projective classification of spin liquids, \cite{2002_wen} this class of {\em Ans\"{a}tze} describe $U(1)$ spin liquids. Preservation of time reversal symmetry suggests that there is at least one gauge in which $\chi_{ij}$ is real and ${\bf E}_{ij}$ is imaginary for all bonds on which they are non zero. 

We choose
\begin{align}
\nonumber
\chi_{ij}&=\chi(>0)\ \ \ \forall i,j\in {\rm nearest\ neighbours}\\
&=0\ \ \ {\rm otherwise}.
\end{align}
For ${\bf E}_{ij}$, there are several choices. We note that, second nearest-neighbours on a pyrochlore lattice belong to two different sublattices (see Fig. \ref{fig_pyrochlore}) which may be thought to be connected through an intermediate atom belonging to a third type of sublattice. Here we discuss two possible {\em Ans\"{a}tze} for ${\bf E}_{ij}$ having the form
\begin{align}
{\bf E}_{ij}=i{\bf E}^{\alpha\beta}=iE\  \hat{\bf n}^{\alpha\gamma\beta}
\end{align} 
where $\alpha,\beta,\gamma=1,2,3,4$ refer to the sublattices (see Fig. {\ref{fig_pyrochlore}}) and $\hat {\bf n}^{\alpha\gamma\beta}$ is an unit vector for the path connecting sublattices $\alpha$ and $\beta$ through the sub-lattice $\gamma$ with $\alpha\neq\beta\neq\gamma$. The direction of the unit vector is chosen to preserve various symmetries of the lattice. We find that there are 4 independent unit vectors $\hat{\bf n}$ and the rest are constrained by symmetry (see Appendix \ref{appen_ansatz}). These two {\em Ans\"{a}tze} are chosen because they break the least number of symmetries of the Hamiltonian. For the first {\em Ansatz} we take\\

{\underline{Ansatz-I} :} 
\begin{align}
\nonumber
\hat{\bf n}^{132}&=\hat{\bf n}^{213}=-\hat{\bf n}^{123}=\hat{\bf x},\ \ 
\hat{\bf n}^{142}=\hat {\bf n}^{214}=-\hat{\bf n}^{124}=\hat{\bf z},\\
\hat{\bf n}^{143}&=\hat{\bf n}^{314}=-\hat{\bf n}^{134}=\hat{\bf z},\ \
\hat{\bf n}^{243}=\hat{\bf n}^{324}=-\hat{\bf n}^{234}=\hat {\bf z}.
\end{align}
Here $\hat{\bf x},\hat{\bf z}$ refers to the unit vectors along the $x$ and $z$ axes respectively. Though translationally invariant, this ansatz breaks some of the point group symmetries of the lattice. An {\em Ansatz} that preserves all the symmetries is a variant of the Kane-Mele construction: {\cite{2009_guo}}

{\underline{Ansatz-II} :}
\begin{align}
\nonumber
\hat{\bf n}^{123}&=-\hat{\bf n}^{132}=-\hat{\bf n}^{213}=\hat{\bf A},\ \ 
\hat{\bf n}^{124}=-\hat {\bf n}^{142}=-\hat{\bf n}^{214}=\hat{\bf B},\\
\hat{\bf n}^{134}&=-\hat{\bf n}^{143}=-\hat{\bf n}^{314}=\hat{\bf C},\ \
\hat{\bf n}^{234}=-\hat{\bf n}^{243}=-\hat{\bf n}^{342}=\hat {\bf D}.
\end{align}
where, $\hat{\bf A}=\frac{1}{\sqrt{3}}\left[1,1,-1\right], \hat{\bf B}=\frac{1}{\sqrt{3}}\left[-1,1,-1\right], \hat{\bf C}=\frac{1}{\sqrt{3}}\left[-1,1,1\right],$ and $\ \ \hat{\bf D}=\frac{1}{\sqrt{3}}\left[1,1,1\right]$ refer to the four $C_3$ axes of the tetrahedra forming the pyrochlore lattice (see Appendix \ref{appen_ansatz}).
\subsection{Emergent spin-orbit coupling and mean-field spinon Hamiltonian}

\begin{figure*}
\centering
\subfigure[]{
\includegraphics[scale=0.3]{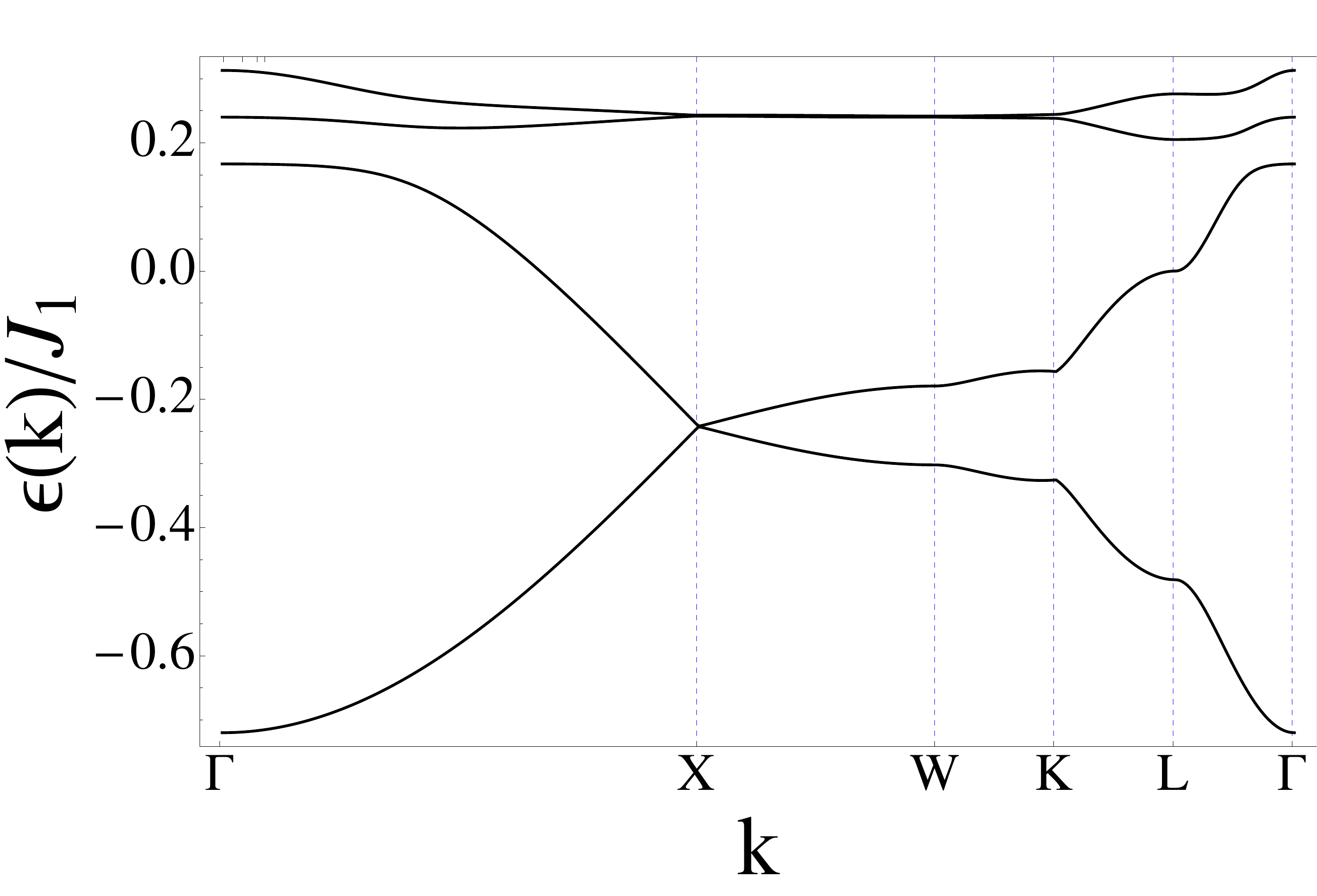}
}
\subfigure[]{
\includegraphics[scale=0.3]{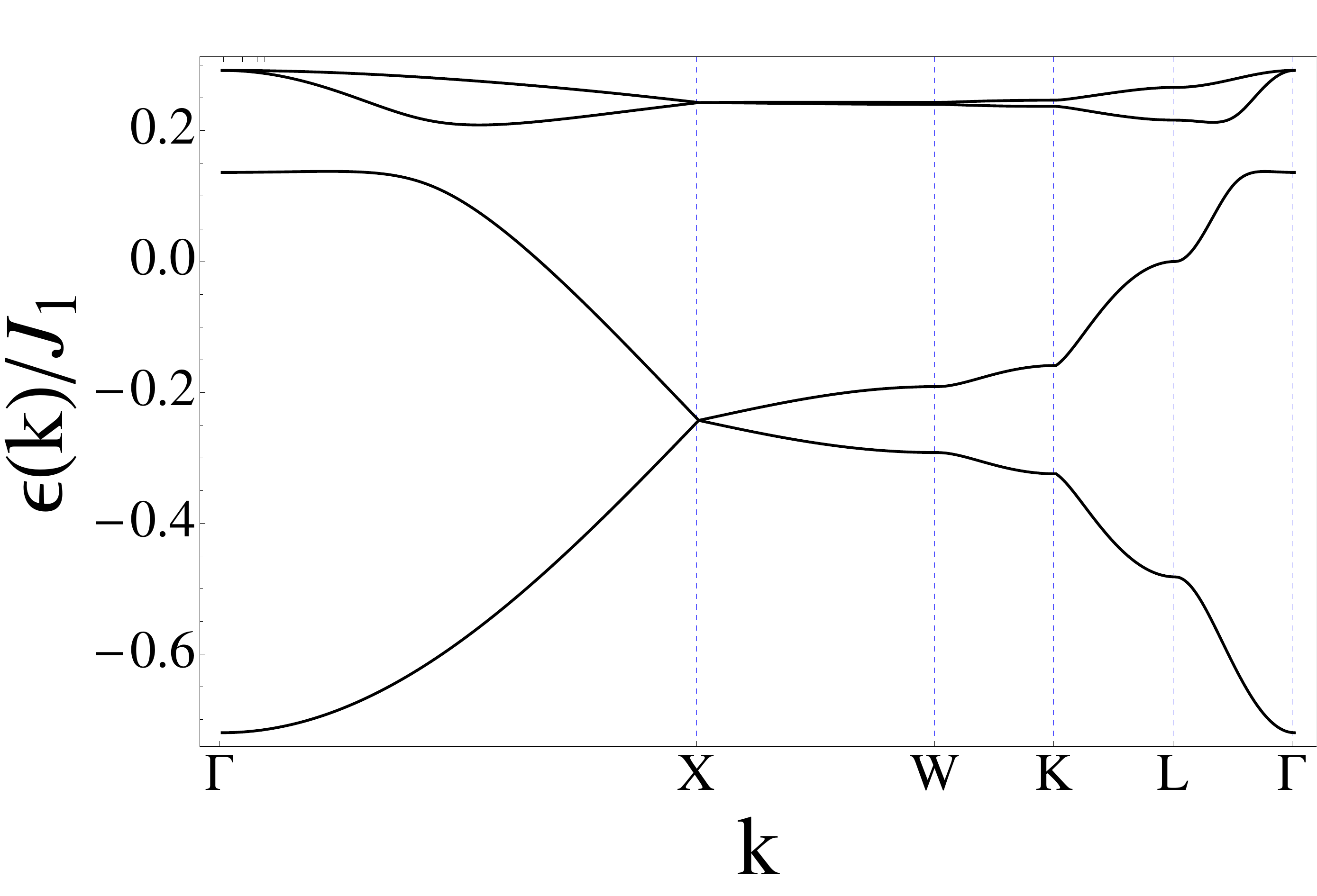}
}
\caption{The spinon band structure is shown along the high-symmetry directions ($J_2/J_1=0.05$) for (a) {\em Ansatz}-I and (b) {\em Ansatz}-II. The values of the mean-field parameters are given in Table \ref{tab_parameters}. Each band is doubly degenerate due to inversion symmetry.}
\label{fig_band}
\end{figure*} 
It is important to note that the above {\em Ans\"{a}tze} give rise to  emergent SO couplings for the spinons even when such interactions are absent for the underlying electrons [Eq. (\ref{eq_hamiltonian}) is spin rotation invariant]. In fact, both the above forms of triplet decoupling spontaneously break the spin $SU(2)$ completely (without breaking time reversal symmetry), which is essential to generate a strong TI in three spatial dimensions. 

More specifically, the quadratic part of the mean field spinon Hamiltonian becomes (details given in the Appendix \ref{appen_spin_liquid})
\begin{align}
H_{MF}=&-\frac{3J_1\chi}{8}\sum_{\langle i\alpha,j\beta\rangle}\sum_{\tau}\left[f^\dagger_{i,\alpha,\tau}f_{j,\beta,\tau}+h.c.\right]\\
\nonumber
&-\frac{i3J_2}{8}\sum_{\langle\langle i\alpha,j\beta\rangle\rangle}\sum_{\tau\tau'}\left[\left(E^{\alpha\beta}_af^\dagger_{i,\alpha,\tau}\sigma^a_{\tau\tau'}f_{j,\beta,\tau'}-h.c\right)\right]
\label{eq_mft}
\end{align}
where as before $\alpha,\beta$ indicate the four sub-lattices, $i,j$ the unit cell, $\tau,\tau'$ the spins, and $a=x,y,z$ axes.

Equation (\ref{eq_mft}) can be diagonalized in the Fourier space and this leads to the spinon Bloch Hamiltonian. The detailed form of the spinon Bloch Hamiltonian and the self consistency of the mean-field parameters are given in Appendix \ref{appen_spin_liquid}. Hence, we get the spinon band structure as a function of ${\bf E}$ and $\chi$ which is shown in Fig. \ref{fig_band} (for self-consistent parameters). The pyrochlore lattice is described by a fcc lattice with a four-point unit cell. So, there are four bands each of which are doubly degenerate due to inversion symmetry (see below). When $\chi\neq 0$ and $E=0$, we have two flat bands lying above two dispersing bands. The flat bands touch the dispersing bands at the center of the Brillouin zone (BZ). With one spinon per site, the two dispersing bands are filled and the spinon Fermi surface reduces to a point at the BZ center. On introducing $E$, an energy gap opens up at the BZ center. Once again the lower two bands are filled and the upper two are empty. This gives us the ``spinon band insulator.'' For $J_2/J_1=0.05$,  the minimum of the mean field energy and the 
corresponding values of the parameters as well as the energy per site, $\mathcal{E}$, are given in Table {\ref{tab_parameters}}. In the table, a constant $-3J_1/16$ per site has been subtracted. The origin of this constant can be traced to the shift in the energy due to the spinon decoupling. Here, we make an estimate of the constant shift by calculating the energy of the nearest-neighbour dimerized state which is $-3 J_1/8$ per site. {\cite{1990_rokshar}} Evaluation of such a constant is useful to compare the energy of this spin liquid state with that of a magnetically ordered state. Both of our {\em Ans\"{a}tze} appear to have similar energies at the mean-field level.

At this point we briefly note that when $J_2 = 0$, the two top bands are flat and touch a dispersing band at the $\Gamma$ point. The flat band, near the $\Gamma$ point, may be looked upon as a quadratically dispersing band in the limit of infinite band mass. Indeed, if we add a small third-neighbour antiferromagnetic coupling, there is quadratic band touching at the $\Gamma$ point. A tree level scaling analysis shows that the scaling dimensions of the fermionic spinon fields at this free fixed point is $3/2$ (with dynamical critical exponent, $z= 2$). The next-nearest-neighbour interaction, when written in terms of spinons, is a short-ranged four-fermion interaction with a coupling that has a tree-level scaling dimension of $-1$. This tree-level scaling argument shows that at the free fixed point the short-ranged four-fermion term is irrelevant in the RG sense. This is in conformity with the irrelevance of short-range four fermion interactions at the free fixed point with quadratic band touching points in three spatial dimensions. This means that we need to introduce finite $J_2$ to get a phase transition to a spinon topological insulator. Notice, however, that our self-consistent mean-field theory, which is quite similar to Hartree-Fock decoupling, shows that the critical value of $J_2$ beyond which it stabilizes the spinon TI is quite small. We also note that in the spin-rotation invariant Hamiltonian, the three-fold degeneracy at the $\Gamma$ point (of two flat bands and one quadratically dispersing band when $J_2 = 0$) is protected by point group symmetry of the pyrochlore lattice (transform as the basis vectors of the so-called $T_2$ irreducible representation). On decoupling the four-fermion term coming from the NNN in the spin triplet channel, the spin rotation symmetry is broken and correct irreducible representations are now obtained from the direct product of the point group and the spinor representation of the magnetic group. The three-fold degeneracy is no longer protected and only a two-fold degeneracy remains (related to the 2-dimensional irreducible representation of the combined symmetry group). This is seen in {\em Ansatz}-II where the point group symmetry is intact while the spin-rotation symmetry is broken (note that each band has a further two-fold degeneracy coming from preservation of inversion symmetry). Thus a band gap is allowed to open up to give a spinon TI.

It is useful to compare the above mean-field energy with that of the magnetically ordered ground state that has been proposed for this system. \cite{1991_reimers} The classical ground state has incommensurate magnetic order with wave vector ${\bf Q}=2\pi(h,h,0)$, where $h\approx 0.741$ for $J_2/J_1=0.05$. \cite{2008_chern} The actual arrangement of the spins is not known. The classical ground-state energy, within spherical approximation (that is consistent with Monte Carlo results \cite{2008_chern}), is $\mathcal{E}_{mag}\approx -0.28\ J_1$ per site for $J_2/J_1=0.05$. While the above estimates coming from mean-field calculations must be taken in a qualitative sense, it shows that the present spin liquid saddle points are energetically competitive with the magnetically ordered ground state. On further comparison with other spin liquid states proposed on the pyrochlore lattice, \cite{2008_burnell} we find that the present state fares quite well within the mean-field approximation. This raises the possibility of stabilizing a spinon TI state in the present as well as related models (e.g., including further neighbours such as $J_3$). However these microscopic details can only be clarified through careful numerical calculations in future.

\begin{table}[h]
\caption{Various parameters at the minimum of the mean-field energy, $\mathcal{E}$, for $J_2/J_1=0.05$.}
\label{tab_parameters}
\begin{tabular}{|c|c|c|c|}\hline
Ansatz & $\chi$ & $E$ & $\mathcal{E}$\\ \hline
I & $0.32$ & $0.52$ & $-0.33\ J_1$ \\ \hline
II& $0.32$ & $-0.6$ & $-0.34\ J_1$ \\ \hline
\end{tabular}
\end{table}

\subsection{The $Z_2$ topological indices}

The spinon band structure allows us to calculate the $Z_2$ topological invariants characterizing it. Such $Z_2$ invariants fully characterize the topological properties of the spinon bands in the presence of time-reversal symmetry (see Ref. \onlinecite{2007_fu} and references therein).

The time-reversal symmetry is given by
\begin{align}
\Theta=(\mathcal{I}_4 \otimes i\sigma^y) K
\end{align}
where $\mathcal{I}_4$ is the four-dimensional identity matrix operating in the sublattice space and $\sigma^y$ is the Pauli matrix operation in the spin space. $K$ is the complex conjugation operator. It is easy to show that the mean-field Hamiltonian is time reversal invariant. 

In addition to this, neither of the  spin liquid ans\"{a}tze break lattice parity. The pyrochlore lattice has inversion symmetry about a site. Taking sub-lattice $1$ as the origin of the unit cell (Fig. \ref{fig_pyrochlore}), we see that the the momentum space representation of the inversion operator is {\cite{2009_guo}}
\begin{align}
\mathcal{P}({\bf q})=(p({\bf q})\otimes \mathcal{I}_2)
\end{align}
where,
\begin{align}
p({\bf q})=\left[\begin{array}{cccc}
1 & 0 & 0 & 0 \\
0 & e^{-i {\bf q}\cdot{\bf a}_1} & 0 & 0\\
0 & 0 & e^{-i {\bf q}\cdot{\bf a}_2} & 0 \\
0 & 0 & 0 & e^{-i {\bf q}\cdot{\bf a}_3} \\
\end{array}\right]
\end{align}
and $p({\bf q})$ operates in the sublattice space and $\mathcal{I}_2$ is the two-dimensional identity that operates in the spin space. ${\bf a}_1, {\bf a}_2$, and ${\bf a}_3$ are the vectors defining the position of the three sublattices with respect to sub-lattice 1 (see Appendix \ref{appen_pyro}). Again it can be shown that the mean-field Hamiltonian is invariant under inversion. 

We therefore wish to calculate the topological invariants for the band structure. For 3D TIs, one strong $(\nu_0)$ and three weak $(\nu_1\nu_2\nu_3)$ $Z_2$ invariants are present. The eight inequivalent time-reversal-invariant momenta (TRIM) within the first Brillouin zone of the fcc lattice are denoted as $\Gamma$ (one), $X$ (three), and $L$ (four) points. Since the system has inversion symmetry we can use the simplified methods of Fu {\em et al.}; {\cite{2007_fu}} i.e., we can calculate the $Z_2$ invariants from the parity eigenvalues at the TRIMs for the filled bands. We note that each band is doubly degenerate. At each TRIM ($i$), a product of the parity eigenvalues is defined as
\begin{align}
\delta_i=\prod_{m=1}^{N}\xi_{2m}(\Gamma_i)
\end{align}
where, $2N$ is the total number of occupied bands and $\xi_{2m}(\Gamma_i)$ denotes the parity at TRIM for one of the degenerate Kramers pairs. We find that for $J_2/J_1=0.05$, at the saddle point for both the ans\"{a}tze considered in this paper, $\delta_i=-1$ at all TRIMs except at $\Gamma$ where it is $+1$.

Once $\delta_i$s are obtained, the strong topological invariant is given by \cite{2007_fu}
\begin{align}
(-1)^{\nu_0}=\prod_{i=1}^8\delta_i
\end{align}
The three weak invariants are given by:
\begin{align}
(-1)^{\nu_k}=\prod_{n_k=1;n_{j\neq k}=0,1}\delta_{i=(n_1n_2n_3)}
\end{align}

Here we have
\begin{align}
\nu_0=1; \ \ \ \nu_1=\nu_2=\nu_3=0
\end{align}

This gives a {\em strong} topological insulator of the class
\begin{align}
(\nu_0;\nu_1\nu_2\nu_3)=(1;000).
\end{align}
An immediate fallout, as noted before, is the presence of robust gapless spinons on all the surfaces. Consider a boundary between the above spin liquid and an ``ordinary" $U(1)$ spin liquid with gapped spinons. Such a surface must have robust gapless spinon surface states. In this sense, this ``symmetry protected topological order" can be regarded as a tool to discover a finer classification (or richer structures) scheme for various kinds of spin liquids. Since the spinons are charge-neutral objects under the external electromagnetic field, they do not carry a charge current. However, such gapless states can carry heat current and hence contribute to the thermal conductivity. Such robust ``metallic'' surface thermal conductivity is one of the signatures of this state. We wish to point out that what happens at the boundary of the above spin liquid with vacuum is a subtle question. Recent calculations on simpler two-dimensional models \cite{2011_kitaev} show that it may depend on the nature of the boundary and hence involve classification of boundary conditions. Extension of these ideas to the present case will require much more sophisticated calculations which is beyond the scope of this paper.

\subsection{Bond nematic order}

The presence of non zero ${\bf E}_{ij}$ (and/or ${\bf D}_{ij}$) leads to bond spin-nematic order characterizing the broken spin rotation symmetry. In particular, Shindou {\em et al.} \cite{2009_shindou} showed that the bond spin-nematic operator, 
\begin{align}
\mathcal{Q}_{ij}^{ab}=\frac{1}{2}\left( S^a_iS^b_j+S^b_jS^a_i\right)-\frac{\delta_{ab}}{3}{\bf S}_i\cdot{\bf S}_j 
\end{align}
gains a non zero expectation value {\cite{2009_shindou}},{\em i.e},
\begin{align}
\langle \mathcal{Q}^{ab}_{ij}\rangle=-\frac{1}{2}\left[E_{ij,a}E_{ij,b}^*-\frac{\delta_{ab}}{3}\vert{\bf E}_{ij}\vert^2\right]\neq 0,
\end{align}
where we have put ${\bf D}_{ij}=0$ and $ij$ are second neighbours. In addition, on incorporating the single spinon per site constraint exactly, we get \cite{2009_shindou}
\begin{align}
{\boldsymbol{\mathcal{J}}}_{ij}=\langle {\bf S}_{i}\times {\bf S}_{j}\rangle =\frac{i}{2}\left[{\bf E}^*_{ij}\chi_{ik}\chi_{kj}-{\bf E}_{ij}\chi_{jk}\chi_{ki}\right]\neq 0,
\end{align}
where $ij$ are second neighbours and $k$ is the intermediate site connecting $i$ and $j$; we have again used ${\bf D}_{ij}=0$. 

Both $\mathcal{Q}_{ij}^{ab}$ and $\boldsymbol{\mathcal{J}}_{ij}$ are even under time reversal and are usually referred as the $n$-{\em nematic} and the $p$-{\em nematic} order parameters, respectively. \cite{2009_shindou,1984_andreev} However, the present phase is different from the conventional bond spin-nematic state since it supports deconfined spinons \cite{1999_balents,2000_senthil,2004_senthil} (In principle, there can be a different spin liquid phase with spin-nematic order where ${\bf E}_{ij}=0$, but, $\mathcal{Q}^{\alpha\beta}_{ij}\neq 0$ and/or $\boldsymbol{\mathcal{J}}_{ij}\neq 0$. A discussion of such phases is beyond the scope of the present mean field scheme).  The order parameter for this spin-nematic, described by the non-collinear vector field ${\bf E}_{ij}$ and uniform singlet field $\chi$,  lives in an $SO(3)$ manifold. Under $\pi$ rotation around the diagonal axis of the hexagonal loops generated by neighbouring tetrahedra in a pyrochlore lattice (Fig. \ref{fig_hex}),
\begin{align}
 {\bf E}_{ij}\rightarrow -{\bf E}_{ij}
 \end{align}
for both the {\em Ans\"{a}tze}. While $\mathcal{Q}^{\alpha\beta}_{ij}$ is even, $\boldsymbol{\mathcal{J}}_{ij}$ is odd under this transformation. The two states described by sets of $\{{\bf E}_{ij}\}$ and $\{-{\bf E}_{ij}\}$ are energetically degenerate. Such a degeneracy will be lifted by small Dzyaloskinshi-Moriya interactions. Further, since $\boldsymbol{\mathcal{J}}_{ij}$ is the local spin current, it couples to the local electric field, $\boldsymbol{\mathscr{E}}$ (this is the physical electric field and not the emergent electric field discussed elsewhere in the paper) through the Aharonov-Casher effect: \cite{1984_aharonov} 
\begin{align}
\epsilon^{abc} \mathcal{J}^a_{i,i+b} \mathscr{E}^c.
\end{align}
 Then, a nonzero value of the order parameter, $\mathcal{J}^a$,  will generate a local electric field and causes a small but finite lattice distortion that can be, in principle, detected. An important question is about the textures of this order parameter and the quantum numbers they carry. Specifically, do they carry electric charge of the emergent gauge field? or, what are their statistics? Details of such issues form interesting future directions.

\section{Elementary excitations, beyond mean-field theory, and outlook}
\label{sec_excitation}

In this last section we discuss the elementary excitations in the spinon TI state and possible effects beyond mean-field theory, as well as phase transitions out of this state. 

The spectrum of low energy excitations in the spinon TI state is quite rich. The Goldstone modes arising from the spin-nematic order, discussed above, are related to the transverse amplitude fluctuations of ${\bf E}_{ij}$, whereas the {\em photon} is related to the phase fluctuation of ${\bf E}_{ij}$ and ${\chi_{ij}}$ about their saddle points.  There is an indirect coupling between the two that can be obtained by integrating out the gapped spinons in the bulk. However, such couplings are inversely proportional to the spinon gap. Also, since the ${\bf E}_{ij}$ does not carry any gauge charge, such couplings, at most, have a dipolar form. The above considerations suggest that such couplings are unimportant in the bulk. At the surface, where the spinons become gapless, the effect of the gauge photon and the Goldstone mode is much more subtle and requires careful consideration. All gapless bosonic modes are expected to contribute a bulk specific heat that scales as $\sim T^3$ at low temperatures. It can be shown that the Goldstone boson couples to the ``spinon-spin current". Recent calculations {\cite{2010_krempa}} indicate that since the bosonic fields live in one dimension higher than the spinon fields, they only have a marginal effect on the spinon self energy. 

In addition to the above excitations, the emergent compact $U(1)$ gauge field also allows a magnetic monopole excitation. Such magnetic monopoles are gapped in the spin liquid and hence unimportant at low energy. On the other hand, when the gapped spinons in the bulk are integrated out, this generates a $\theta$-term, $\left(\theta/2\pi\right){\bf e}\cdot{\bf b}$, where ${\bf e}$ and ${\bf b}$ are the emergent ``electric'' and ``magnetic'' fields and $\theta=\pi$ for topological insulators. \cite{2010_moore,2009_qi,2008_qi,2009_essin,2010_vazifeh} It is known that in the presence of such a $\theta$-term, the magnetic monopoles acquire ``electric" charges and become {\em dyons} \cite{1979_witten,2010_rosenberg}. It is interesting to think about phase transitions out of the present nematic spin liquids by condensing these dyons. Such a transition serves as a potential example of {\em oblique confinement}. \cite{1981_thooft} Since the dyons carry both electric and magnetic charges, their condensation may lead to the Meissner effect for both the ${\bf e}$ and ${\bf b}$ fields. The ${\bf e}$ and ${\bf b}$ fields are odd under parity and time reversal, respectively. In the case of bosonic spinons coupled to compact $U(1)$ gauge field (and the $\theta$-term absent), the theory of the ``electric" confinement in the presence of background ``electric" charges (spinons) is well understood \cite{1991_read,1989_read,2005_motrunich}. In that case, the ``electric" confinement due to monopole condensation leads to the breaking of lattice parity and the resultant state has valence bond solid ordering. In the light of these known results, it is tempting to speculate on the fate of the state obtained by condensing the dyons in the present case. It is easy to show that in the presence of the $\theta$-term, a finite expectation value of ${\bf e}$ implies the same for ${\bf b}$. Both $\langle {\bf e}\rangle,\langle {\bf b}\rangle\neq 0$ implies a state that breaks both time reversal and parity. A possible candidate is one where both valence bond ordering and magnetic orders are present. More exotic possibilities include states with a non-zero spin chirality coexisting with valence bond order. A transition to such a state from the nematic spin liquid, considered in this paper, if continuous, is forbidden within the conventional Landau-Ginzburg-Wilson paradigm, and would represent a new universality class of quantum phase transitions.

\acknowledgements
YBK and DHL acknowledge the generous support of the Aspen Center for Physics, where this work was initiated. We are grateful to S. Kivelson for useful discussions and a critical reading of the manuscript. We also thank F. Burnell, T. Dodds, W.W-Krempa, Y. Ran, S. S. Ray, R. Schaffer and V. S. Venkataraman for fruitful discussions. This work was supported by the NSERC (SB, YBK, SSL), CIFAR, CRC (SB,YBK) and DOE Grant No. DE-AC02-05CH11231 (DHL). SSL acknowledges the support from ERA of the Ontario Ministry of Research and Innovation. Research at the Perimeter Institute is supported in part by the Government of Canada through Industry Canada, and by the Province of Ontario through the Ministry of Research and Information.

\appendix
\section{The Pyrochlore Lattice}
\label{appen_pyro}
To describe the pyrochlore lattice, we take the conventional cubic unit cell and measure distances in units of its sides {\cite{2010_conlon}}. The pyrochlore lattice is then described by a fcc lattice with 4-point basis (one tetrahedron at each site of the fcc lattice; see Fig. \ref{fig_pyrochlore}).  The basis vectors are
\begin{align}
{\bf a}_1=\frac{1}{2}\left(\bf{\hat z} +\bf{\hat y}\right), \  \  {\bf a}_2=\frac{1}{2}\left(\bf{\hat z} +\bf{\hat x}\right), \  \
{\bf a}_3=\frac{1}{2}\left(\bf{\hat x} +\bf{\hat y}\right).
\end{align}
The reciprocal lattice vectors are then given by
\begin{align}
\nonumber
{\bf b}_1=2\pi & \left({\bf \hat y} +\bf{\hat z} -\bf{\hat x}\right), \  \  {\bf b}_2=2\pi\left(\bf{\hat z} +\bf{\hat x}-{\bf \hat y}\right),\\
&{\bf b}_3=2\pi\left(\bf{\hat x} +\bf{\hat y}-{\bf \hat z}\right).
\end{align}

The 4-point basis may be taken as
\begin{align}
{\bf{C}}_\mu=\left\{0,\frac{{\bf a}_1}{2},\frac{{\bf a}_2}{2},\frac{{\bf a}_3}{2}\right\}
\end{align}
where $\mu=1,2,3,4$ denotes the 4 sublattices. The 6 bond vectors are now defined as.
\begin{align}
{\bf{d}}_{\mu\nu}={\bf{C}}_\mu-{\bf{C}}_\nu
\end{align}
\section{Constraints on the spin liquid ans\"{a}tze}
\label{appen_ansatz}

As pointed out in the main text, the second neighbour connections may be usefully thought of as being mediated through an intermediate atom. On a pyrochlore lattice all these three belong to different sublattices; e.g., sites belonging to sublattices $1$ and $2$ are connected through atoms belonging to sub-lattices $3$ and $4$, and so forth. We denote such paths as 
\begin{align}
\alpha\rightarrow\gamma\rightarrow\beta).
\end{align}
 Now, in a pyrochlore lattice, the tetrahedra form hexagonal loops. Each such hexagon has sites belonging to any three kinds of sub-lattices. Thus there are four kinds of hexagons containing the following participating sub-lattices:
\begin{align}
(1,2,3);\ \ \ (1,2,4);\ \ \ (1,3,4);\ \ \ (2,3,4).
\end{align}

\begin{figure}
\centering
\includegraphics[scale=0.725]{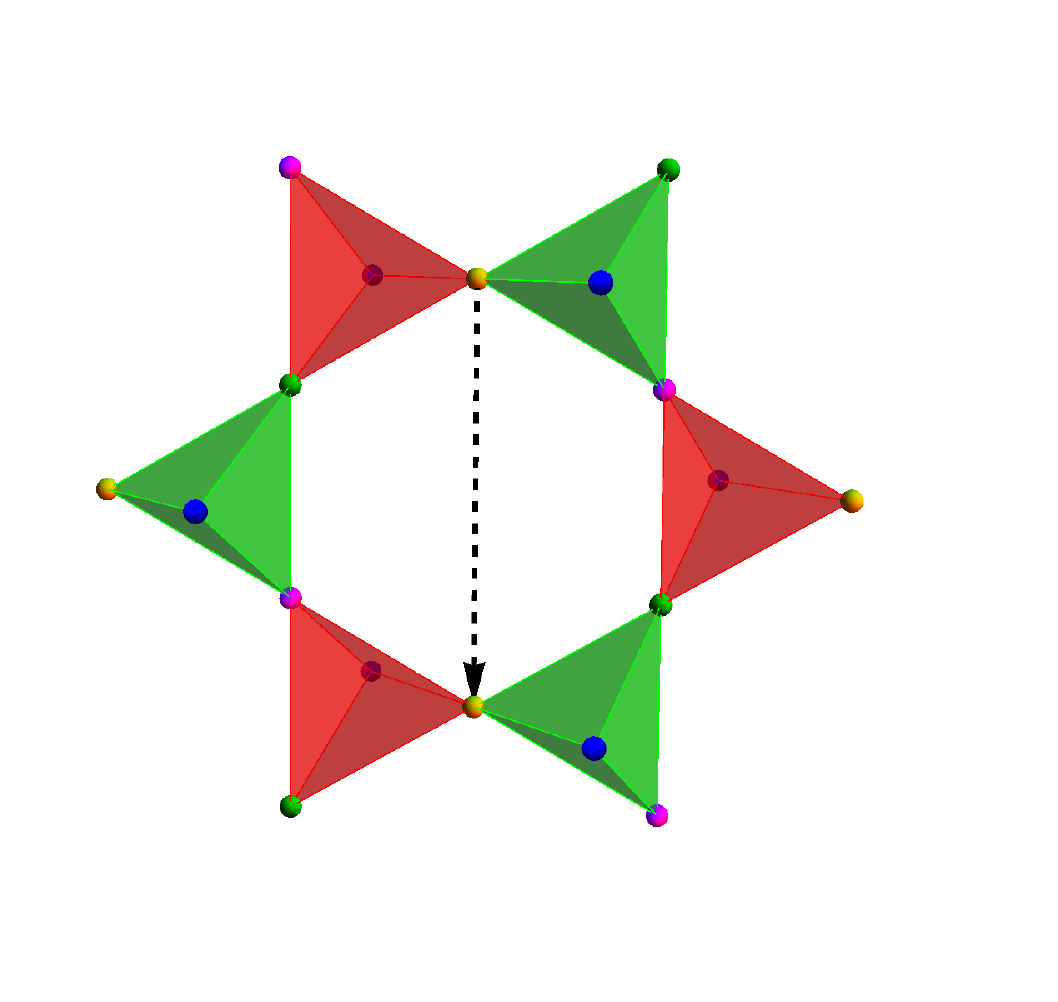}
\caption{A $\pi$-rotation axis about the hexagon (formed by the neighbouring tetrahedra in green and red) is shown by a dashed line.}
\label{fig_hex}
\end{figure}

There is a $\pi$-rotational symmetry about an axis joining two similar (same sub-lattice) atoms belonging to each hexagon. An example is shown in figure {\ref{fig_hex}}. This means that there are only four different allowed ${\bf E}_{ij}$ vectors. These may be denoted by

\begin{align}
{\bf A}, {\bf B}, {\bf C}, {\bf D}.
\end{align}
Once we choose four vectors, the others are completely specified as shown below. If we choose (say)
\begin{align}
\nonumber
(1\rightarrow 2\rightarrow 3)&: {\bf A}\\
\nonumber
(1\rightarrow 2\rightarrow 4)&: {\bf B}\\
\nonumber
(1\rightarrow 3\rightarrow 4)&: { \bf C}\\
(2\rightarrow 3\rightarrow 4)&: {\bf D},
\end{align}
then, we must have (due to the $\pi$-rotation symmetry)
\begin{align}
\nonumber
(1\rightarrow 3\rightarrow 2)&: {-\bf A};\ \ \ (2\rightarrow 1\rightarrow 3): {-\bf A},\\
\nonumber
(1\rightarrow 4\rightarrow 2)&: {-\bf B};\ \ \ (2\rightarrow 1\rightarrow 4): {-\bf B},\\
\nonumber
(1\rightarrow 4\rightarrow 3)&: {-\bf C}\ \ \ (3\rightarrow 1\rightarrow 4): {-\bf C},\\
(2\rightarrow 4\rightarrow 3)&: {-\bf D};\ \ \ (3\rightarrow 2\rightarrow 4): {-\bf D}.
\end{align}
The other paths are set by Hermitian conjugation. We can show that this choice is translationally invariant under the translation of the fcc lattice. 

We now find further constraints on these four vectors, as set by the different transformations of the point group $T_d$. To this end we list the different elements of the point group. They are:
\begin{align}
T_d: E(1),\{c_3,c_3^2\}(8),\{s_4,s_4^3\}(6),s_2(3),\sigma_d(6)
\end{align}
These are the 24 elements which can be divided in to 5 classes. As a consequence there are five irreducible representations: 2 one-dimensional, 1 two-dimensional, and 2 three-dimensional. It is enough to see the transformation of the four above vectors under the four $c_3$ rotations. These represent three fold rotations about the vertices of the tetrahedron. The transformations are given by the following:
\begin{enumerate}

\item Under $c_3$ through sublattice $1$.
\begin{align}
\nonumber
& 1\rightarrow 1,\ \ 2\rightarrow 3,\ \ 3\rightarrow 4,\ \ 4\rightarrow 2\\ 
&\Rightarrow {\bf A}\rightarrow {\bf C},\ \ {\bf B}\rightarrow {-\bf A},\ \ {\bf C}\rightarrow {-\bf B}, \ \ {\bf D}\rightarrow {\bf D}
\end{align}

\item Under $c_3$ through axis 2 we have:
\begin{align}
\nonumber
& 2\rightarrow 2,\ \ 1\rightarrow 4,\ \ 3\rightarrow 1,\ \ 4\rightarrow 3\\
& \Rightarrow {\bf A}\rightarrow {-\bf B},\ \ {\bf B}\rightarrow {\bf D},\ \ {\bf D}\rightarrow {-\bf A}, \ \ {\bf C}\rightarrow {\bf C}
\end{align}

\item Under $c_3$ through axis 3 we have:
\begin{align}
\nonumber
& 3\rightarrow 3,\ \ 1\rightarrow 2,\ \ 2\rightarrow 4,\ \ 4\rightarrow 1\\
&\Rightarrow {\bf A}\rightarrow {-\bf D},\ \ {\bf C}\rightarrow {\bf A},\ \ {\bf D}\rightarrow {-\bf C}, \ \ {\bf B}\rightarrow {\bf B}
\end{align}

\item Under $c_3$ through axis 4 we have:
\begin{align}
\nonumber
&4\rightarrow 4,\ \ 1\rightarrow 3,\ \ 2\rightarrow 1,\ \ 3\rightarrow 2\\
&\Rightarrow {\bf B}\rightarrow {-\bf C},\ \ {\bf C}\rightarrow {-\bf D},\ \ {\bf D}\rightarrow {\bf B}, \ \ {\bf A}\rightarrow {\bf A}
\end{align}
\end{enumerate}

Similarly, we can consider the other transformations. However it is easy to see from the above transformations that a characteristic feature of the four vectors is that one of them remain invariant under the three-fold rotation. This suggests that we can choose (up to a sign) the four vectors as the axis of rotation for the four threefold rotation axes. These then satisfy all the point group symmetries as well as the translation symmetry. This is exactly the form for {\em Ansatz} II. On the other hand, it is now easy to see why {\em Ansatz} I violates the point group symmetries.

\section{The mean field Spin liquid state}
\label{appen_spin_liquid}

In this subsection, for our convenience, we split the Hamiltonian in Eq. (\ref{eq_hamiltonian}) as
\begin{align}
H=H_1+H_2
\end{align}
where, $H_1$ and $H_2$ refer to the nearest- and second-neighbour parts, respectively. We consider a $U(1)$ spin liquid by setting the pairing terms [both singlet and triplet in Eqs. (\ref{eq_fm_decouple}), (\ref{eq_afm_decouple})] to zero. We shall also drop the constant terms. However we note that such constants will be important when we want to compare the energies of the spin liquid and the magnetically ordered state.
\subsubsection{Nearest Neighbour AFM exchange}
For the AFM decoupling we decouple in the singlet channel and have:

\begin{align}
H_1=-\frac{3J_1}{8}\sum_{\langle ij\rangle}\left[\vert \chi_{ij}\vert^2-\left(\chi^*_{ij}  f^{\dagger}_{i\alpha}f_{j\alpha}+h.c.\right)\right].
\end{align}
In our spin liquid ansatz, we consider uniform $\chi_{ij}=\chi$ and noting that each spin has 6 neighbours, we get ($N$ is the total number of spins)
\begin{align}
\frac{H_1}{J_1 N}=\frac{9\chi^2}{8}-\frac{3\chi}{8N}\sum_{\langle ij\rangle}\left[f^\dagger_{i\sigma}f_{j\sigma}+h.c.\right].
\end{align}
Now we introduce the 4 sub-lattice fcc lattice and use the Fourier transform:
\begin{align}
f_{i,\mu,\sigma}=\frac{1}{\sqrt{N_{FCC}}}\sum_{{\bf{q}}\in BZ} f_{{\bf q},\mu,\sigma}r^{i{\bf q}\cdot({\bf r}_i+{\bf C}_\mu)}\ \ \ \ \ (N_{FCC}=N/4)
\end{align}
where ${\bf q}\in BZ$ denotes the summation over the Brillouin zone of the fcc lattice. This gives:
\begin{align}
\frac{H_1}{J_1 N}=\frac{9\chi^2}{8}-\frac{3\chi}{32N_{FCC}}\sum_{{\bf{q}}\in BZ}\Psi_{{\bf q}}H^{(1)}{({\bf q})}\Psi_{{\bf q}}
\end{align}
where $\Psi_{{\bf q}}=\left[f_{{\bf q}1\uparrow},f_{{\bf q}2\uparrow},f_{{\bf q}3\uparrow},f_{{\bf q}4\uparrow},f_{{\bf q}1\downarrow},f_{{\bf q}2\downarrow},f_{{\bf q}3\downarrow},f_{{\bf q}4\downarrow}\right]^T$ and $H^{(1)}_{\bf q}$ equals
\begin{align}
H^{(1)}({\bf q})=\left[\begin{array}{cc}
\mathcal{H}^{(1)}({\bf q}) & 0\\
0 & \mathcal{H}^{(1)}({\bf q})\\
\end{array}\right]
\end{align}
where $\mathcal{H}^{(1)}({\bf q})$ is given by
\begin{align}
2\left[\begin{array}{cccc}
0 & \cos {\left[\frac{q_y+q_z}{4}\right]}& \cos {\left[\frac{q_x+q_z}{4}\right]}& \cos {\left[\frac{q_x+q_y}{4}\right]} \\
\cos {\left[\frac{q_y+q_z}{4}\right]} & 0 & \cos {\left[\frac{q_x-q_y}{4}\right]}& \cos {\left[\frac{q_x-q_z}{4}\right]} \\
\cos {\left[\frac{q_x+q_z}{4}\right]} & \cos {\left[\frac{q_x-q_z}{4}\right]} & 0& \cos {\left[\frac{q_y-q_z}{4}\right]}\\
\cos {\left[\frac{q_x+q_y}{4}\right]} & \cos {\left[\frac{q_x-q_z}{4}\right]} & \cos {\left[\frac{q_y-q_z}{4}\right]}&0\\
\end{array}\right]
\end{align}
\subsubsection{Second-neighbour FM exchange}

By using the already chosen ansatz ${\bf E}_{ij}$(noting that there are 12 second neighbours), we get,
\begin{align}
\nonumber
\frac{H_2}{J_1 N}&=\left(\frac{J_2}{J_1}\right)\left[\frac{9 E^2}{4}-\frac{3E}{8} \frac{1}{N}\sum_{{\bf q}\in BZ} \Psi_{{\bf q}}^\dagger H^{(2)}({\bf q})\Psi_{{\bf q}}\right]\\ 
&=\left(\frac{J_2}{J_1}\right)\left[\frac{9 E^2}{4}-\frac{3E}{32} \frac{1}{N_{FCC}}\sum_{{\bf q}\in BZ} \Psi_{{\bf q}}^\dagger H^{(2)}({\bf q})\Psi_{{\bf q}}\right]
\end{align}
where
\begin{align}
H^{(2)}({\bf q})=\left[\begin{array}{cc}
\mathcal{A}({\bf q}) & \mathcal{B}({\bf q})\\
\mathcal{B}^\dagger({\bf q}) & - \mathcal{A}({\bf q})\\
\end{array}\right]
\end{align}
$\mathcal{A}({\bf q})$ and $\mathcal{B}({\bf q})$ are $4\times 4$ matrices whose forms differ for {\em Ans\"{a}tze} I and II. 
\subsection{The band structure and self-consistency of the mean field parameters}
Within our {\em Ansatz}, the mean-field Hamiltonian is
\begin{align}
\nonumber
\frac{H}{J_1 N}&=\left[\frac{9\chi^2}{8}+\left(\frac{J_2}{J_1}\right)\frac{9 E^2}{4}\right]\\
\nonumber
&\ -\frac{1}{4N_{FCC}}\sum_{{\bf{q}}\in BZ}\Psi^\dagger_{{\bf q}}\left[\frac{3\chi}{8}H^{(1)}({\bf q})+\left(\frac{J_2}{J_1}\right)\frac{3E}{8}H^{(2)}({\bf q})\right]\Psi_{{\bf q}}\\
&=\left[\frac{9\chi^2}{8}+\left(\frac{J_2}{J_1}\right)\frac{9 E^2}{4}\right]+\frac{1}{4N_{FCC}}\left[\sum_{{\bf{q}}\in BZ}\Psi^\dagger_{{\bf q}}H({\bf q})\Psi_{{\bf q}}\right]
\end{align}
where, in the last expression we have used
\begin{align}
H({\bf q})=-\left[\frac{3\chi}{8}H^{(1)}({\bf q})+\left(\frac{J_2}{J_1}\right)\frac{3E}{8}H^{(2)}({\bf q})\right].
\end{align}
These are $8\times 8$ matrices. Their eigenvalues form four doubly degenerate bands. With one spinon per site, the two lowest bands are filled. Let the dispersion of the two low lying bands be $\lambda_1({\bf q})$ and $\lambda_2({\bf q})$. Therefore the mean-field ground-state energy of the spin liquid is given by
\begin{align}
\nonumber
\frac{\mathcal{E}}{J_1 N}=&\left[\frac{9\chi^2}{8}+\left(\frac{J_2}{J_1}\right)\frac{9 E^2}{4}\right]\\
&+\frac{1}{4N_{FCC}}\sum_{{\bf q}\in BZ}\left(2\lambda_1({\bf q})+2\lambda_2({\bf q})\right)
\end{align} 
The factor $2$ multiplying the bands comes from the fact that each band is doubly degenerate. The minimum of the mean-field energy and the corresponding values of $\chi$ and $E$ are given in Table \ref{tab_parameters}.


\end{document}